\documentclass[runningheads,a4paper]{llncs}

\usepackage{booktabs}   %% For formal tables:
\usepackage{subcaption} %% For complex figures with subfigures/subcaptions
\usepackage{latexsym}
\usepackage{amssymb}
\usepackage{amsmath}
\usepackage{stmaryrd}
\usepackage[all]{xy}
\usepackage{graphicx}
\usepackage{hyperref}
\usepackage{multicol}

\newcommand{\Log}{\mathcal{W}}

\newcommand{\ignore}[1]{}

\newcommand{\arro}[1]{\xrightarrow{#1}} 
\newcommand{\hoo}{\hookrightarrow}

\newcommand{\comp}{\:|\:}

\newcommand{\h}{\mathit{h}}

\renewcommand{\k}{\lambda}
\renewcommand{\k}{\ell}

\newcommand{\lh}{\leftharpoondown}
\newcommand{\rh}{\rightharpoonup}

\newcommand{\sql}{\mbox{$\lfloor\hspace{-.5ex}\lfloor$}}
\newcommand{\sqr}{\mbox{$\rfloor\hspace{-.4ex}\rfloor$}}

\usepackage{xcolor}

\newcommand{\red}[1]{{\color{red} #1}}

\newcommand{\blue}[1]{{\color{blue} #1}}

\newcommand{\purple}[1]{{\color{purple} #1}}

\newcommand{\id}{id}
\newcommand{\ol}[1]{\overline{#1}}  % sequence of objects

\newcommand{\dom}{{\mathcal{D}om}}

\newcommand{\nil}{[\:]}

\newcommand{\trace}{\mathcal{L}}

\newcommand{\sender}{\mathit{sender}}
\newcommand{\parent}{\mathit{parent}}

\newcommand{\consl}{\!:\!}
\newcommand{\cons}{\mbox{$+$}}

\def \tuple#1{\langle #1 \rangle}

\long\def\comment#1{}

\begin{document}

\title{Causal-Consistent Reversible Debugging:\\ Improving 
CauDEr%
\thanks{This work has been partially supported by EU (FEDER) and
    Spanish MCI/AEI under grants TIN2016-76843-C4-1-R and
    PID2019-104735RB-C41, by the \emph{Generalitat Valenciana}
    under grant Prometeo/2019/098 (DeepTrust), and by
    French ANR project DCore ANR-18-CE25-0007.}
}

\author{Juan Jos\'e Gonz\'alez-Abril  \and Germ\'an Vidal}

\institute{
MiST, VRAIN, Universitat Polit\`ecnica de Val\`encia\\
\email{juagona6@posgrado.upv.es,gvidal@dsic.upv.es}
}

\maketitle

\begin{abstract}
	Causal-consistent reversible debugging allows one to 
	explore concurrent computations 
	back and forth in order to locate the source of an error.
	In this setting, backward steps can be chosen freely as long
	as they are \emph{causal consistent}, i.e., as long as all the actions 
	that depend on the action we want to undo
	 have been already undone.
	Here, we consider a  framework for 
	causal-consistent reversible debugging in the
	functional and concurrent language Erlang. This framework
	considered programs translated to an intermediate
	representation, called Core Erlang. Although using such
	an intermediate representation simplified both the formal
	definitions and their implementation in a debugging tool,
	the choice of Core Erlang also complicated the use of the debugger.
	In this paper, we extend the framework in order to deal with 
	\emph{source} Erlang programs,
	also including some features that were not considered before.
	Moreover, we integrate the two existing approaches (user-driven 
	debugging and replay debugging)
	into a single, more general framework, 
	and develop a new version of the debugging tool
	\textsf{CauDEr} including all the mentioned extensions
	as well as a renovated user interface.\\[2ex]
  Published as \emph{Gonz\'alez-Abril, J.J. and Vidal, G. (2021). Causal-Consistent Reversible Debugging: Improving CauDEr. 
  In: Morales, J.F.\ and Orchard, D.A.\ (eds) Practical Aspects of Declarative Languages. PADL 2021. Lecture Notes in Computer Science, vol 12548. Springer.}\\[2ex]  
  The final authenticated publication is available at
   \url{https://doi.org/10.1007/978-3-030-67438-0\_9}
\end{abstract}

%\keywords{concurrency, logging, causal-consistent
%  debugging}

%%%%%%%%%%%%%%%%%%%%%%%%%%%%%%%%%%%%%%%%%%%%%%%%%%%%%%%%%%%%%%%%%%%
\section{Introduction} \label{intro}

Reversible debugging is a well established technique 
\cite{Gri70,Zel73} which advocates that, in order to find the
location of an error,  it is often more natural
to explore a computation \emph{backwards} from the 
observable misbehavior. Therefore, reversible debuggers allow one
to explore a computation back and forth.
There are already a number of 
software debuggers that follow this idea 
(e.g., Undo \cite{UndoWhitePaper}). Typically, 
one can undo the steps of a computation in exactly the
inverse order of the original forward computation.
However, in the context of a concurrent language like Erlang, 
reversible debugging
becomes much more complex. On the one hand, in these 
languages, there is no clear (total) order for the actions of
the concurrent processes, since the semantics is often
nondeterministic. On the other hand, 
one is typically interested in a particular process,
e.g., the one that exhibited a misbehavior. Thus,
undoing the steps of \emph{all} processes in the
very same order of the forward computation is very inconvenient, 
since we had to go through many actions 
that are completely unrelated 
to the process of interest. 

Recently, \emph{causal-consistent} 
reversible debugging \cite{GLM14} has been
introduced in order to overcome these problems. In this setting,
concurrent actions can be undone freely as long as they are
causal-consistent, i.e., no action is undone until all the actions 
that depend on this one have been already undone. For instance,
one cannot undo the spawning of a  process until all
the actions of this process have been undone. 

A reversible semantics for (Core) Erlang was first introduced in
\cite{NPV16b} and, then, extended and improved in \cite{LNPV18jlamp}.
A causal-consistent reversible debugger, \textsf{CauDEr}, that follows these
ideas was then presented in \cite{LNPV18}.\footnote{To the best
of our knowledge, CauDEr is the first \emph{causal-consistent}
reversible
debugger for a realistic programming language. Previous approaches
did not consider concurrency, only allowed a \emph{deterministic} 
replay---e.g., the case of rr \cite{OCJFHNP17} and 
ocamldebug \cite{ocaml}---or considered a very simple language,
as in \cite{GLM14}. The reader is referred to \cite{LPV19} for a 
detailed comparison of CauDEr with other, related approaches.
} 
In the original approach,
both the forward and the backward computations were driven by
the user (i.e., the user selects the process of interest as well as 
the message
to be received when there are several possibilities), 
so we refer to this approach as \emph{user-driven} debugging.
However, when a computation fails, it is often very
difficult, even impossible, to replicate the same computation
in the debugger. This is a well-known problem in the debugging of
concurrent programs. Therefore, \cite{LPV19} introduces a 
novel approach, called \emph{replay} debugging, 
that is based on a light program instrumentation so that
the execution of the instrumented program 
produces an associated \emph{log}.
This log can then be used for the debugger 
to replay the same computation 
or a \emph{causally equivalent} one (i.e., one that at least
respects the same order between dependent actions).

Unfortunately, all these works consider the intermediate
language Core Erlang \cite{Car01}. Dealing with (a subset of)
Core Erlang
made the theoretical developments easier (without loss of
generality, since programs are automatically transformed from
source Erlang to Core Erlang during the compilation process).
The reversible debugger \textsf{CauDEr} considers Core Erlang too,
which greatly simplified the implementation task.
Unfortunately, as a result, the debugger is rather
difficult to use since the programmer is often not familiar with the
Core Erlang representation of her program. 
Compare, for instance, the following Erlang code defining
the factorial function:
\begin{verbatim}
  fact(0)          -> 1;
  fact(N) when N>0 -> N * fact(N-1).
\end{verbatim}
and the corresponding representation in Core Erlang:
{\small 
\begin{verbatim}
'fact'/1 =
  fun (_0) ->
    case _0 of
      <0> when 'true' -> 1
      <N> when call 'erlang':'>'(N, 0) ->
          let <_1> = call 'erlang':'-'(N, 1)
          in let <_2> = apply 'fact'/1(_1)
          in call 'erlang':'*'(N, _2)
      <_3> when 'true' ->
          primop 'match_fail'({'function_clause',_3})
    end
end
\end{verbatim}}
\noindent
On the other hand, while Erlang remains
relatively stable, the Core Erlang specification changes
(often undocumented) with some releases of the Erlang/OTP
compiler, which makes maintaining the debugger quite a
difficult task.

In this paper, we extend the causal-consistent approach to
reversible debugging  of Erlang in order to deal with 
(a significant subset of) the source language. 
The main contributions are the following:
\begin{itemize}
\item We redefine both the standard and the reversible
semantics in order to deal with source Erlang expressions. 
\item Moreover, we integrate the two previous approaches, the 
\emph{user-driven} reversible debugger of \cite{LNPV18jlamp} and
the \emph{replay} debugger of \cite{LPV19}, intro a single,
more general framework.
\item Finally, we have implemented a new version of the 
\textsf{CauDEr} reversible debugger \cite{LNPV18} which implements
the above contributions, also redesigning the user interface.
\end{itemize}
Missing details can be found in the appendix.

%%%%%%%%%%%%%%%%%%%%%%%%%%%%%%%%%%%%%%%%%%%%%%%%%%%%%%%%%%%%%%%%%%% 
\section{The Source Language} \label{sec:erlang}

In this section, we informally introduce the syntax and semantics
of the considered language, a significant subset of Erlang. 
A complete, more formal 
definition can be found in the appendix. We also
discuss the main differences with previous work that considered
Core Erlang instead.

Erlang is a typical higher-order, eager functional language
extended with some additional features for message-passing
concurrency. Let us first consider the sequential component
of the language, which includes the following elements:
\begin{itemize}
\item \textit{Variables}, denoted with identifiers that start with an 
uppercase letter.

\item \textit{Atoms}, denoted with identifiers that start with a lowercase
letter. Atoms are used, e.g., to denote constants and function names.

\item \textit{Data constructors}.  Erlang only considers 
lists (using Prolog notation, i.e., $[e_1|e_2]$ denotes a list with
first element $e_1$ and tail $e_2$) and tuples, denoted by an
expression of the form $\{e_1,\ldots,e_n\}$, where $e_1,\ldots,e_n$
are expressions, $n\geq 0$. However, Erlang does not allow
user-defined constructors as in, e.g., Haskell.

\item \textit{Values}, which are built from literals (e.g., numbers),
atoms and data constructors.

\item \textit{Patterns}, which are similar to values but might
also include variables.

\item \textit{Expressions}. Besides variables, values and patterns, 
we consider the following types of expressions:
\begin{itemize}
 \item \emph{Sequences} of the form $e_1,\ldots,e_n$ where $e_i$ is
 a single expression (i.e., it cannot be a sequence), $i=1,\ldots,n$.
 Evaluation proceeds from left to right: first, we evaluate $e_1$ to
 some value $v_1$ thus producing $v_1,e_2,\ldots,e_n$. If $n>1$,
  the sequence is then reduced to $e_2,\ldots,e_n$, and so forth. 
  Therefore, sequences are eventually evaluated to a (single) value.
 The use of sequences of expressions in the right-hand sides of 
 functions is a distinctive feature 
 of Erlang. In the following, we assume that all expressions are
 single expressions except otherwise stated.
 
 \item \emph{Pattern matching} equations of the form 
 $p = e$, where $p$
 is a pattern and $e$ is an expression. Here, we evaluate $e$ to
 a value $v$ and, then, try to find a substitution $\sigma$ such that
 $p\sigma=v$. If so, the equation is reduced to $v$ and $\sigma$
 is applied to the complete expression. E.g., the expression 
 ``$\tt \{X,Y\} = \{ok,40+2\},X$" is reduced to ``$\tt \{ok,42\},ok$"
 and, then, to $\tt ok$.
 
\item  \emph{Case} expressions like $\mathsf{case}~e
~\mathsf{of}~cl_1;\ldots;cl_n~\mathsf{end}$, 
where each clause $cl_i$
has the form ``$p_i ~[\mathsf{when}~g_i]\to e_i$" with
$p_i$ a pattern, $g_i$ a \emph{guard} and
$e_i$ an expression (possibly a sequence), $i=1,\ldots,n$.
\emph{Guards} are optional, and 
can only contain calls to built-in functions (typically, arithmetic
and relational operators). A case expression then selects the
first clause such that the pattern matching equation $p_i=e$
holds for some substitution $\sigma$ and $g_i\sigma$ reduces 
to $\mathsf{true}$. Then, the case expression reduces to
$e_i\sigma$.

\item \emph{If} expressions have the form $\mathsf{if}~g_1\to e_1;
\ldots; g_n\to e_n~\mathsf{end}$, where $g_i$ is a guard and
$e_i$ is an expression (possibly a sequence),
$i=1,\ldots,n$. It proceeds in the obvious way by returning the first 
expression $e_i$ such that the corresponding guard $g_i$
holds.

\item \emph{Anonymous functions} (often called \emph{funs} 
in the language Erlang) have the form 
``$\mathsf{fun}~(p_1,\ldots,p_n) ~[\mathsf{when}~g]
\to e~\mathsf{end}$,\!"
where $p_1,\ldots,p_n$ are patterns, $g$ is a guard (optional), 
and $e$ is an expression (possibly a sequence).

\item \emph{Function calls} have the usual form, $f(e_1,\ldots,e_n)$,
where $f$ is either an atom (denoting a function name) or a 
fun, and $e_1,\ldots,e_n$ are expressions.
\end{itemize} 
\end{itemize}
Concurrency in Erlang mainly follows the \emph{actor model},
where processes (actors) interact through message sending and
receiving. Here, we use the term \emph{system} to refer to the
complete runtime application. In this scheme, each process has 
an associated \emph{pid} (for \textit{p}rocess \textit{id}entifier) 
that is unique in the system. Moreover, processes are supposed
to have a local mailbox (a queue) where messages are stored
when they are received, and until they are \emph{consumed}. 
In order to model concurrency, the following elements are
introduced:
\begin{itemize}
\item The built-in function $\mathsf{spawn}$ is used to create 
a new process. Here, for simplicity, we assume that the arguments
are a function name and a list of arguments. E.g., the expression
$\mathsf{spawn}(\mathit{foo},[e_1,e_2])$ evaluates $e_1,e_2$ to some
values $v_1,v_2$, spawns a new process with a fresh pid $p$
that evaluates $\mathit{foo}(v_1,v_2)$ as a side-effect, and returns 
the pid $p$.

\item Message sending is denoted with an expression of the
form $e_1 \:!\: e_2$. Here, expressions $e_1,e_2$ are first 
evaluated to some values $v_1,v_2$ and $v_2$ is returned.
Moreover, as a side effect, 
$v_2$ (the \emph{message}) is eventually stored in the 
mailbox of the process with pid $v_1$ (if any).

\item Messages are consumed with a 
statement of the form 
$\mathsf{receive}~cl_1;\ldots;cl_n~\mathsf{end}$.
The evaluation of a receive statement is similar to a case
expression of the form 
$\mathsf{case}~v~\mathsf{of}~cl_1;\ldots;cl_n~\mathsf{end}$,
where $v$ is the first message in the process' mailbox
that matches some clause. When no message in
the mailbox matches any clause (or the mailbox is empty),
computation \emph{suspends} until a matching message arrives.
Then, as a side effect, the selected message is removed 
from the process' mailbox.

\item Finally, the ($0$-ary) \emph{built-in} function $\mathsf{self}$ 
evaluates to the pid of the current process.
\end{itemize}
An Erlang program is given by a set of function definitions, where
each function definition has the form
\[
\begin{array}{lllll}
f(p_{11},\ldots,p_{1n}) & [\mathsf{when} & g_1] & \to & e_1;\\
f(p_{21},\ldots,p_{2n}) & [\mathsf{when} & g_2] & \to & e_2;\\
\ldots\\
f(p_{m1},\ldots,p_{mn}) & [\mathsf{when} & g_m] & \to & e_m.
\end{array}
\]
Where $p_{ij}$ is a pattern, $g_i$ is a guard (optional), and
$e_i$ is an expression (possibly a sequence), $i=1,\ldots,m$.
Let us illustrate the main ingredients of the language with a 
simple example:

\begin{example}
Consider the program shown in Figure~\ref{fig:stock}. 
Here, we consider that the execution starts with the call
$\tt main()$. Function $\tt main$ then spawns two new
processes that will evaluate $\tt customer1(S)$
and $\tt customer2(S)$, respectively, where $\tt S$ is
the pid of the current process (the \emph{server}).
Finally, it calls $\tt server(0)$, where function $\tt server$
implements a simple server to update the current stock
(initialized to $0$). It accepts three types of requests:

\begin{figure}[t]
\begin{multicols}{2}
\begin{verbatim}
main() ->
  spawn(customer1, [self()]),
  spawn(customer2, [self()]),
  server(0).

server(N) ->
  receive
    {add,M} 
        -> server(N+M);
    {del,M,C} when N>=M 
        -> K = N-M, C ! K, server(K);
    stop 
        -> ok
  end.

customer1(S) ->
  S ! {add,3},
  S ! {del,10,self()},
  receive
    N -> io:format("Stock: ~p~n",[N])
  end,
  S ! stop.

customer2(S) ->
  S ! {add,5},
  S ! {add,1},
  S ! {add,4}.
\end{verbatim}
\end{multicols}
\caption{A simple Erlang program.}\label{fig:stock}
\end{figure}

\begin{itemize}
\item $\tt \{add,M\}$: in this case, it simply calls the 
server with the updated argument $\tt N+M$.\footnote{Note that
the \emph{state} of the process is represented by the argument
of the call to function $\tt server$.}

\item $\tt \{del,M,C\}$: assuming that $\tt N>=M$ holds,
the server computes the new stock ($\tt N-M$), sends it
back to the customer and, then, calls the server with the 
updated value.

\item Finally, $\tt stop$ simply terminates the execution of
the server.
\end{itemize}
Each customer performs
a number of requests to the server, also waiting
for a reply when the message is $\tt \{del,10,self()\}$ since
the server replies with the updated stock in this case.
Here, $\tt format$ is a \emph{built-in} function with the usual
meaning, that belongs to module $\tt io$.\footnote{In Erlang,
function calls are often prefixed by the module where the
function is defined.}

Note that we cannot make any assumption regarding the
order in which the messages from these two customers 
reach the server. In particular, if message $\tt \{del,10,self()\}$
arrives when the stock is smaller than $\tt 10$, it will stay
in the process' mailbox until all the messages  from 
$\tt customer2$ 
are received (i.e., $\tt  \{add,5\}$, $\tt \{add,1\}$, and
$\tt \{add,4\}$).
\end{example}
Let us now consider the semantics of the language.
In some previous formalization \cite{LNPV18jlamp}, 
a system included both
a \emph{global mailbox}, common to all processes, and
a \emph{local mailbox} associated to each process.
Here, when a message is sent, it is first stored in the global
mailbox (which is called the \emph{ether} in \cite{SFB10}).
Then, eventually, the message is delivered to the target
process and stored in its local mailbox, so that it can be
consumed with a receive statement. In this paper,
similarly to \cite{LPV19}, we abstract away the local
mailboxes and just consider a single global mailbox.
There is no loss of generality in this decision, since one can
just define an appropriate structure of queues in the
global mailbox so that it includes all local mailboxes of
the system. Nevertheless, for simplicity, we will assume
in the following that our global mailbox is just a set of
messages of the form $\{p,p',v\}$, where $p$ is the
pid of the sender, $p'$ is the pid of the target, and $v$
is the message. We note that this abstraction has no
impact 
for \emph{replay} debugging since one follows the steps
of an actual execution anyway. For user-driven debugging it
might involve exploring some computations that are not
feasible though.

In the following, a \emph{process} is denoted by a 
configuration of the form $\tuple{p,e,\theta,S}$, where
$p$ is the process' pid, $e$ is an expression (to be
evaluated), $\theta$ is a substitution (the current 
environment), and $S$ is a stack (initially empty, see below).
A \emph{system} is then denoted by $\Gamma;\Pi$, 
where $\Gamma$ is a global mailbox and 
$\Pi$ is a pool of processes, denoted as 
$ \tuple{p_1,\theta_1,e_1,S_1} ~\comp \cdots
\comp~\tuple{p_n,\theta_n,e_n,S_n} $; here ``$\comp$'' 
represents an associative and commutative operator.
We often denote a system as
$ \Gamma; \tuple{p,\theta,e,S}\comp\Pi $ to point out that
$\tuple{p,\theta,e,S}$ is an arbitrary process of the pool
(thanks to the fact that ``$\comp$'' is associative and
commutative).

An \emph{initial system} has the form
$\{\:\};\tuple{p,\id,e,\nil}$, where $\{\:\}$ is an empty global mailbox,
$p$ is a pid, $\id$ is the identity substitution, $e$ is an
expression (typically a function application that starts the
execution), and $\nil$ is an empty stack.

Our (reduction) semantics is defined at two levels: first, we 
have a  (labelled) transition relation on expressions. Here, we
define a typical higher-order, eager small-step semantics
for sequential expressions. In contrast to previous approaches
(e.g., \cite{LNPV18jlamp,LPV19}), we introduce the use of
stacks in order
to avoid producing illegal expressions in some cases.
Consider, for instance, the following function definition:
\[
\tt foo(X) \to Y=1, X+Y.
\]
and the expression ``$\tt \mathsf{case}~foo(41)~\mathsf{of}~
R \to R~\mathsf{end}$.\!" 
Here, by unfolding the call to function $\tt foo$ we might
get ``$\tt \mathsf{case}~Y=1,42+Y~\mathsf{of}~
R \to R~\mathsf{end}$,\!" 
which is not legal since sequences of expressions
are not allowed in the argument of a case expression.
A similar situation might occur when evaluating a case or an
if expression, since they can also return a sequence of
expressions. We avoid all these illegal intermediate expressions 
by moving the current environment to a
stack and starting a subcomputation. When the subcomputation
ends, we recover the environment from the stack and continue with
the original computation. E.g., the following rules
define the evaluation of a function call:
\[
\begin{array}{l}
    (\mathit{Call1}) ~ {\displaystyle
    \frac{\mathsf{match\_fun}((v_1,\ldots,v_n),\mathsf{def}(f/n,\mathsf{P})) = (\sigma,e)}{\red{\theta},\red{C[}f(v_1,\ldots,v_n)\red{]},S
        \arro{\tau} \sigma,e,\red{(\theta,C[\_])}\consl S}} %\{\ol{X_n\mapsto v_n}\},e}}
    \\[4ex]

        (\mathit{Return}) ~ {\displaystyle
      \frac{}{\sigma,v,\red{(\theta,C[\_])}\consl S
        \arro{\tau} \red{\theta},\red{C[}v\red{]},S}} %\{\ol{X_n\mapsto v_n}\},e}} 
\end{array}
\]
Here, $C[e]$ denotes an arbitrary (possibly empty)
evaluation  \emph{context} 
where $e$ is the next expression to be reduced according to
an eager semantics. The auxiliary functions $\mathsf{def}$ and
$\mathsf{match\_fun}$ are used to look for the definition of
a function $f/n$ in a program $\mathsf{P}$ 
and for computing the corresponding matching 
substitution, respectively; here, $e$ is the body
of the selected clause and $\sigma$ is the matching substitution.
If  a value is eventually obtained, rule \textit{Return} applies and
recovers the old environment   $(\theta,C[\_])$
from the stack.

Regarding the semantics of expressions with side effects
(spawn, sending and receiving messages, and self), we label
the step with enough information for the next level---the
system semantics---to perform the side effect. For instance,
For spawning a process, we have these two rules:
\[
\begin{array}{l}
     (\mathit{SpawnExp}) ~
      %\multicolumn{1}{c}
      {\displaystyle
       \frac{}{\theta,C[\mathsf{spawn}(f,[\ol{v_n}])],S
         \arro{\mathsf{spawn}(\kappa,f,[v_1,\ldots,v_n])} \theta,C[\kappa],S         
       }}
       \\[3ex]

      (\mathit{Spawn}) \\
       \multicolumn{1}{c}{\displaystyle
        \frac{\theta,e,S \arro{\mathsf{spawn}(\kappa,f,[v_1,\ldots,v_n])}
          \theta',e',S'~~\mbox{and}~~ p'~\mbox{is a fresh pid}}{\Gamma;\tuple{p,\theta,e,S} 
          \comp \Pi %%\;\instarro{\red{\{p,\mathsf{spawn}(p')\}}}\; 
           \hoo %_{\red{p,\mathsf{spawn}(p')}}
                          \Gamma;\tuple{p,\theta',e'\{\kappa\mapsto
                          p'\},S'}\comp \red{\tuple{p',\id,f(v_1,\ldots,v_n),\nil}} 
          \comp \Pi}
      }
\end{array}
\]
Here, the first rule just reduces a call to 
spawn to a fresh variable $\kappa$,
a sort of ``future", since the pid of the new process is not
visible at the expression level.  The step is  labelled with 
$\mathsf{spawn}(\kappa,f,[v_1,\ldots,v_n])$. Then, the system rule
\textit{Spawn} completes the step by adding a new process
initialized to $\tuple{p',\id,f(v_1,\ldots,v_n),\nil}$; moreover, 
$\kappa$ is bound to the (fresh) pid of the new process.
We have similar rules for evaluating the sending and receiving 
of messages, for sequential expressions, etc.
The complete transition rules of both the semantics of expressions
($\to$) and the semantics of systems ($\hoo$) can be found
in the appendix.

The main advantage of this hierarchical definition of the semantics
is that one can produce different \emph{non-standard} versions of the
semantics by only replacing the transition rules for systems. 
For instance, a \emph{tracing semantics} can be simply obtained 
by instrumenting the standard semantics as follows:
\begin{itemize}
\item First, we tag messages with a fresh label, so that we can 
easily relate messages sent and received. Without the labels,
messages with the same value could not be distinguished.
In particular, messages in the global mailbox have now the
form $\{p,p',\{v,\ell\}\}$ instead of $\{p,p',v\}$, where $\ell$
is a label that must be unique in the system.

\item Then, each step $s_1 \hoo_{p,r} s_2$ is labeled 
with a pair $p,r$ where
$p$ is the pid of the selected process and $r$ is 
either $\mathsf{seq}$ for sequential steps, $\mathsf{send}(\ell)$
for sending a message labeled with $\ell$, $\mathsf{rec}(\ell)$
for receiving a message labeled with $\ell$, $\mathsf{spawn}(p)$
for spawning a process with pid $p$, and $\mathsf{self}$
for evaluating a call of the form $\mathsf{self}()$.
\end{itemize}
The complete \emph{tracing semantics} can also be found in 
the appendix. 

As in \cite{LPV19}, we can instantiate to our setting the well-known
\emph{happened-before} relation~\cite{Lam78}. In the following,
we refer to one-step reductions $s \hoo_{p,r} s'$ as \emph{transitions},
and to longer reductions as \emph{derivations}.

\begin{definition}[happened-before, independence] \label{def:causal} Given a derivation $d$ and two transitions
  $t_1 = (s_1 \hoo_{p_1,r_1} s'_1)$ and
  $t_2 = (s_2 \hoo_{p_2,r_2} s'_2)$ in $d$, we say that $t_1$ happened before
  $t_2$, in symbols $t_1\leadsto t_2$, if one of the following
  conditions holds:
  \begin{itemize}
  \item they consider the same process, i.e., $p_1 = p_2$, and $t_1$ comes before $t_2$;
  \item $t_1$ spawns a process $p$, i.e., $r_1 = \mathsf{spawn}(p)$, and
    $t_2$ is performed by process $p$, i.e., $p_2=p$;
  \item $t_1$ sends a message $\k$, i.e., $r_1=\mathsf{send}(\k)$, and
    $t_2$ receives the same message $\k$, i.e.,
    $r_2=\mathsf{rec}(\k)$.
  \end{itemize}
  Furthermore, if $t_1\leadsto t_2$ and $t_2\leadsto t_3$, then
  $t_1\leadsto t_3$ (transitivity). 
  Two transitions $t_1$ and $t_2$ are \emph{independent} if $t_1
  \not\leadsto t_2$ and $t_2\not\leadsto t_1$.
\end{definition}
An interesting property of our semantics is that consecutive
independent transitions can be switched without changing the final
state.

The happened-before relation gives rise to an equivalence relation
equating all derivations that only differ in the switch of independent
transitions. Formally,

\begin{definition}[causally equivalent derivations] \label{def:causally-equivalent}
  Let $d_1$ and $d_2$ be derivations under the tracing semantics. We say that $d_1$ and $d_2$ are
  \emph{causally equivalent}, in symbols $d_1\approx d_2$, if $d_1$ can be
  obtained from $d_2$ by a finite number of switches of pairs of consecutive
  independent transitions.
\end{definition}
The tracing semantics can be used as a model to instrument a
program so that it produces a log of the computation as a side-effect.
This log can then be used to \emph{replay} this computation 
(or a causally equivalent one) in the debugger. Formally,
a \emph{process log} $\omega$ is a (finite) sequence of events 
$(r_1,r_2,\ldots)$ where each $r_i$ is either
$\mathsf{spawn}(p)$, $\mathsf{send}(\k)$ or $\mathsf{rec}(\k)$,
with $p$ a pid and $\k$ a message identifier. A \emph{system log}
$\Log$ is  defined as a partial mapping from pids to 
processes' logs (an empty log is denoted by $\nil$). Here, 
the notation $\Log[p\mapsto \omega]$ is used to denote that
$\omega$ is the log of the process with pid $p$; as usual,
we use this notation either as a condition on a system log 
$\Log$ or as a modification of $\Log$.

Besides defining a tracing semantics that produces a system log
of a computation as a side effect, we can also define a reversible
semantics by instrumenting the rules of the system semantics 
as shown in the next section.

%%%%%%%%%%%%%%%%%%%%%%%%%%%%%%%%%%%%%%%%%%%%%%%%%%%%%%%%
\section{A Causal-Consistent Reversible Semantics} \label{sec:uncontrolled}

In this section, we first present an instrumented semantics
which is reversible, i.e., we define an appropriate 
\emph{Landauer embedding} \cite{Lan61}
for the standard semantics. Then, we introduce a backward 
semantics that proceeds in the opposite direction. 
Both the forward and backward semantics
are \emph{uncontrolled}, i.e., they have several sources
of nondeterminism:
\begin{enumerate}
\item \emph{Direction}: they can proceed both forward and backward.
\item \emph{Choice of process}: in general, several processes
may perform
a reduction step, and an arbitrary one is chosen.
\item \emph{Message delivery}: when there are 
several (matching) messages targeted
to the same process, any of them can be received.
\item Finally, (fresh) pids and message labels are chosen in
a random way.
\end{enumerate}
We note that, when we proceed in \emph{replay} mode, the last
choices (3-4) are made deterministic. Nevertheless, the calculus
is still highly nondeterministic. Therefore, we will finally introduce
a \emph{controlled} version of the semantics where reductions
are driven by the user requests (e.g.,``go ahead until the sending of 
a message labeled with $\ell$", ``go backwards up to the step 
immediately before process $p$ was spawned", etc). The controlled
semantics is formalized as a third layer (on top of  the rules for
expressions and systems).

%%%%%%%%%%%%%%%%%%%%%%%%%%%%%%%%%%%%%%%
\subsection{A Reversible Semantics}

In the following, a \emph{system} is denoted by a triple
$\Log;\Gamma;\Pi$, where $\Log$ is a 
(possibly empty) \emph{system log},
$\Gamma$ is a global mailbox, and $\Pi$ is a pool of processes.
Furthermore, a \emph{process} is now represented by
a configuration of the form $\tuple{p,\h,\theta,e,S}$,
where $p$ is the pid of the process, $\h$ is a process
\emph{history},  $\theta$ is an environment, $e$ is an expression 
to be evaluated, and  $S$ is an stack.
In this context, a history $\h$
records the intermediate states of a process using terms headed by
constructors $\mathsf{seq}$, $\mathsf{send}$, $\mathsf{rec}$,
$\mathsf{spawn}$, and $\mathsf{self}$, and whose arguments are the
information required to (deterministically) undo the step, following a
typical Landauer embedding \cite{Lan61}. 

\begin{figure}[p]
  $
  \hspace{-5ex}\begin{array}{r@{~}c}
    \multicolumn{1}{l}{(\mathit{Seq})} & {\displaystyle
      \frac{\theta,e,S \arro{\tau} \theta',e',S'
      }{\red{\Log};\Gamma;\tuple{p,\blue{\h},\theta,e,S}\comp \Pi
                     \rh_{p,\mathsf{seq},\purple{\{\mathsf{s}\}}}
      \red{\Log};\Gamma;\tuple{p,\blue{\mathsf{seq}(\theta,e,S) \cons \h},\theta',e',S'}\comp \Pi}
      }\\[3ex]

    \multicolumn{1}{l}{(\mathit{Send})} &
   {\displaystyle
      \frac{\theta,e,S \arro{\mathsf{send}(p',v)}
        \theta',e',S' ~~\mbox{and $\red{\ell}$ is a fresh identifier}}{\begin{array}{l@{~}l}
                      \red{\Log[p\mapsto\nil]};\Gamma;\tuple{p,\blue{\h},\theta,e,S} 
        \comp \Pi  \rh_{p,\mathsf{send}(\k), \purple{\{\mathsf{s},\k^\Uparrow\}}} %%\instarrow{\red{\{p,\mathsf{send}(\k)\}}}
            & \red{\Log};\Gamma\cup\{(p,p',\red{\{v,\k\}})\};\\ %%\ins{\Gamma}{p}{p'}{\red{\{v,\k\}}};\\
            &
              \tuple{p,\blue{\mathsf{send}(\theta,e,S,p',\{v,\k\})\cons\h},\theta',e',S'}\comp \Pi
                    \end{array}}
      }\\[7ex]

   & {\displaystyle
      \frac{\theta,e,S \arro{\mathsf{send}(p',v)}
        \theta',e',S'}{\begin{array}{l@{~}l}
                      \red{\Log[p\mapsto\mathsf{send}(\k)\cons \omega]};\Gamma;\tuple{p,\blue{\h},\theta,e,S} 
        \comp \Pi  \rh_{p,\mathsf{send}(\k), \purple{\{\mathsf{s},\k^\Uparrow\}}} %%\instarrow{\red{\{p,\mathsf{send}(\k)\}}}
            & \red{\Log[p\mapsto \omega]};\Gamma\cup\{(p,p',\red{\{v,\k\}})\};\\ %%\ins{\Gamma}{p}{p'}{\red{\{v,\k\}}};\\
            & \tuple{p,\blue{\mathsf{send}(\theta,e,S,p',\{v,\k\})\cons
           \h},\theta',e',S'}\comp \Pi
                    \end{array}}
      }\\[3ex] 

  \multicolumn{1}{l}{(\mathit{Receive})} &
   {\displaystyle
        \frac{\theta,e,S \arro{\mathsf{rec}(\kappa,\ol{cl_n})}
          \theta',e',S'~~\mbox{and}~~ \mathsf{match\_rec}(\ol{cl_n}\theta,v) =
         (\theta_i,e_i)}
                           {\begin{array}{ll}
%%\ext{\Gamma}{p'}{p}{\red{\{v,\k\}}};
\red{\Log[p\mapsto\nil]};\Gamma\cup\{(p',p,\red{\{v,\k\}})\}
\tuple{p,\blue{\h},\theta,e,S}\comp \Pi \\
\hspace{10ex}\rh_{p,\mathsf{rec}(\k), \purple{\{\mathsf{s},\k^\Downarrow\}}} %%\instarrow{\red{\{p,\mathsf{rec}(\k)\}}}
          \red{\Log};\Gamma;\tuple{p,\blue{\mathsf{rec}(\theta,e,S,p',\{v,\k\})\cons\h},\theta'\theta_i,e'\{\kappa\mapsto
            e_i\},S'}\comp \Pi
\end{array}}
      }\\[7ex]

   & {\displaystyle
        \frac{\theta,e,S \arro{\mathsf{rec}(\kappa,\ol{cl_n})}
          \theta',e',S'~~\mbox{and}~~ \mathsf{matchrec}(\theta,\ol{cl_n},v) =
         (\theta_i,e_i)}
                           {\begin{array}{ll}
%%\ext{\Gamma}{p'}{p}{\red{\{v,\k\}}};
\red{\Log[p\mapsto \mathsf{rec}(\k)\cons \omega]};\Gamma\cup\{(p',p,\red{\{v,\k\}})\}
\tuple{p,\blue{\h},\theta,e,S}\comp \Pi \\
\hspace{10ex}\rh_{p,\mathsf{rec}(\k), \purple{\{\mathsf{s},\k^\Downarrow\}}} %%\instarrow{\red{\{p,\mathsf{rec}(\k)\}}}
          \red{\Log[p\mapsto \omega]};\Gamma;\tuple{p,\blue{\mathsf{rec}(\theta,e,S,p',\{v,\k\})\cons\h},\theta'\theta_i,e'\{\kappa\mapsto
            e_i\},S'}\comp \Pi
\end{array}}
      }\\[6ex]

      \multicolumn{1}{l}{(\mathit{Spawn})} &
{\displaystyle
        \frac{\theta,e,S \arro{\mathsf{spawn}(\kappa,mod, f,[\ol{v_n}])}
          \theta',e',S'~~~\mbox{and $\red{p'}$ is a fresh identifier}}{\begin{array}{l@{~}l}
\red{\Log[p\mapsto\nil]};\Gamma;\tuple{p,\blue{\h},\theta,e,S} 
          \comp \Pi \rh_{p,\mathsf{spawn}(p'), \purple{\{\mathsf{s},\mathsf{sp}_{p'}\}}}  %\instarrow{\red{\{p,\mathsf{spawn}(p')\}}}
        & \red{\Log};\Gamma;\tuple{p,
        \blue{\mathsf{spawn}(\theta,e,S,p')\cons\h},\theta',e'\{\kappa\mapsto \red{p'}\},S'}\\
        & \comp \tuple{\red{p'},\blue{\nil},\id,mod\consl f(\ol{v_n}),\nil} \comp \Pi
      \end{array}}
      }\\[7ex]

 & {\displaystyle
        \frac{\theta,e,S \arro{\mathsf{spawn}(\kappa,mod,f,[\ol{v_n}])}
          \theta',e',S'}
        % ~~~\mbox{and}~~~\red{\omega''=\plog(\omega,p')}}
                 {\begin{array}{l@{~}l}
\red{\Log[p\mapsto\mathsf{spawn}(p')\cons\omega]};\Gamma;\tuple{p,\blue{\h},\theta,e,S} 
          \comp \Pi \\
\hfill \rh_{p,\mathsf{spawn}(p'), \purple{\{\mathsf{s},\mathsf{sp}_{p'}\}}}  %\instarrow{\red{\{p,\mathsf{spawn}(p')\}}}
        & \red{\Log[p\mapsto\omega]};\Gamma;\tuple{p,
        \blue{\mathsf{spawn}(\theta,e,S,p')\cons\h},\theta',e'\{\kappa\mapsto \red{p'}\},S'}\\
        & \comp \tuple{\red{p'},\blue{\nil},\id,mod\consl f(\ol{v_n}),\nil} \comp \Pi
      \end{array}}
      }\\[10ex]

%      \multicolumn{1}{l}{(\mathit{Spawn2})} \\ 
%{\displaystyle
%        \frac{\theta,e,S \arro{\mathsf{spawn}(\kappa,\mathit{exprs})}
%          \theta',e',S'~~~\mbox{and $\red{p'}$ is a fresh identifier}}{\begin{array}{l@{~}l}
%\red{\Log[p\mapsto\nil]};\Gamma;\tuple{p,\blue{\h},\theta,e,S} 
%          \comp \Pi \rh_{p,\mathsf{spawn}(p'), \purple{\{\mathsf{s},\mathsf{sp}_{p'}\}}}  %\instarrow{\red{\{p,\mathsf{spawn}(p')\}}}
%        & \red{\Log};\Gamma;\tuple{p,
%        \blue{\mathsf{spawn}(\theta,e,S,p')\cons\h},\theta',e'\{\kappa\mapsto \red{p'}\},S'}\\
%        & \comp \tuple{\red{p'},\blue{\nil},\id, \mathit{exprs},\nil} \comp \Pi
%      \end{array}}
%      }\\[7ex]
%
% {\displaystyle
%        \frac{\theta,e,S \arro{\mathsf{spawn}(\kappa,mod,f,[\ol{v_n}])}
%          \theta',e',S'}
%        % ~~~\mbox{and}~~~\red{\omega''=\plog(\omega,p')}}
%                 {\begin{array}{l@{~}l}
%\red{\Log[p\mapsto\mathsf{spawn}(p')\cons\omega]};\Gamma;\tuple{p,\blue{\h},\theta,e,S} 
%          \comp \Pi \\
%\hfill \rh_{p,\mathsf{spawn}(p'), \purple{\{\mathsf{s},\mathsf{sp}_{p'}\}}}  %\instarrow{\red{\{p,\mathsf{spawn}(p')\}}}
%        & \red{\Log[p\mapsto\omega]};\Gamma;\tuple{p,
%        \blue{\mathsf{spawn}(\theta,e,S,p')\cons\h},\theta',e'\{\kappa\mapsto \red{p'}\},S'}\\
%        & \comp \tuple{\red{p'},\blue{\nil},\id, \mathit{exprs},\nil} \comp \Pi
%      \end{array}}
%      }\\[5ex]

    \multicolumn{1}{l}{(\mathit{Self})} & {\displaystyle
      \frac{\theta,e,S \arro{\mathsf{self}(\kappa)} \theta',e',S'}{\red{\Log};\Gamma;\tuple{p,\blue{\h},\theta,e,S} 
        \comp \Pi \rh_{p,\mathsf{self}, \purple{\{\mathsf{s}\}}} \red{\Log};\Gamma;\tuple{p,\blue{\mathsf{self}(\theta,e,S)\cons\h},\theta',e'\{\kappa\mapsto p\},S'} 
        \comp \Pi }
      }
  \end{array}
  $ 
  \caption{Uncontrolled forward
    semantics} \label{fig:combined-replay-rules-uncontrolled} 
\end{figure}

\begin{figure}[p]
  $
  \begin{array}{r@{~}c}
    (\mathit{\ol{Seq}}) & {\displaystyle
     \begin{array}{l}
     \red{\Log};\Gamma;\tuple{p,\blue{\mathsf{seq}(\theta,e,S) \cons \h},\theta',e',S'}\comp
     \Pi \;\lh_{p,\mathsf{seq}, \purple{\{\mathsf{s}\}\cup \mathcal{V}}}\;
      \red{\Log};\Gamma;\tuple{p,\blue{\h},\theta,e,S}\comp \Pi\\
        \hfill\mbox{where}~\mathcal{V} =
       \dom(\theta')\backslash\dom(\theta)
       \end{array}
      }
      \\[2ex]

    (\mathit{\ol{Send}}) & {\displaystyle
      \begin{array}{l@{~}l}
      %%\ins{\Gamma}{p}{p'}{\red{\{v,\k\}}};
      \red{\Log[p\mapsto\omega]};\Gamma\cup\{(p,p',\red{\{v,\k\}})\};
           \tuple{p,\blue{\mathsf{send}(\theta,e,S,p',\{v,\k\})\cons
           \h},\theta',e',S'}\comp \Pi\\
           \hspace{20ex} \lh_{p,\mathsf{send}(\k), \purple{\{\mathsf{s},\k^\Uparrow\}}} %%\linstarrow{\red{\{p,\mathsf{send}(\k)\}}}
          \red{\Log[p\mapsto \mathsf{send}(\k)\cons\omega]};\Gamma;\tuple{p,\blue{\h},\theta,e,S} 
        \comp \Pi \end{array}
      }\\[3ex]

      (\mathit{\ol{Receive}}) & {\displaystyle
        \begin{array}{ll}
\red{\Log[p\mapsto\omega]};\Gamma;\tuple{p,\red{\omega},\blue{\mathsf{rec}(\theta,e,S,p',\{v,\k\})\cons\h},\theta',e',S'}\comp \Pi\\
\hspace{10ex}  \lh_{p,\mathsf{rec}(\k), \purple{\{\mathsf{s},\k^\Downarrow\}\cup\mathcal{V}}} %%\linstarrow{\red{\{p,\mathsf{rec}(\k)\}}} 
          %%\ext{\Gamma}{p'}{p}{\red{\{v,\k\}}};
          \red{\Log[p\mapsto \mathsf{rec}(\k)\cons\omega]};\Gamma\cup\{(p',p,\red{\{v,\k\}})\};
\tuple{p,\blue{\h},\theta,e,S}\comp
          \Pi \\         
        \hfill\mbox{where}~\mathcal{V} =
       \dom(\theta')\backslash\dom(\theta)
\end{array}
      }\\[4ex] 
      
      (\mathit{\ol{Spawn}}) & {\displaystyle
        \begin{array}{l@{~}l}
          \red{\Log[p\mapsto\omega]};\Gamma;\tuple{p,
        \blue{\mathsf{spawn}(\theta,e,S,p')\cons\h},\theta',e',S'}
        \comp \tuple{\red{p'},\red{\omega'},\blue{\nil},\id,e''} \comp \Pi\\
       \hspace{20ex} \lh_{p,\mathsf{spawn}(p'), \purple{\{\mathsf{s},\mathsf{sp}_{p'}\}}} %%\linstarrow{\red{\{p,\mathsf{spawn}(p')\}}}
       \red{\Log[p\mapsto \mathsf{spawn}(p')\cons\omega]};\Gamma;\tuple{p,\mathsf{spawn}(p')\cons,\blue{\h},\theta,e,S} 
          \comp \Pi  
      \end{array}
      }\\[3ex]

    (\mathit{\ol{Self}}) & {\displaystyle
      \red{\Log};\Gamma;\tuple{p,\blue{\mathsf{self}(\theta,e,S)\cons\h},\theta',e',S'} 
        \comp \Pi \lh_{p,\mathsf{self}, \purple{\{\mathsf{s}\}}} \red{\Log};\Gamma;\tuple{p,\blue{\h},\theta,e,S} 
        \comp \Pi 
      }
  \end{array}
  $ 
  \caption{Uncontrolled backward semantics} \label{fig:combined-rollback-rules-uncontrolled}
\end{figure}

The rules of the (forward) reversible semantics are shown in 
Figure~\ref{fig:combined-replay-rules-uncontrolled}. The subscripts
of the arrows can be ignored for now. They will become relevant 
for the controlled semantics.
The premises of the rules consider the reduction of an expression,
$\theta,e,S  \arro{\mathit{label}} \theta',e',S'$, where the
\textit{label} includes enough information to perform the
corresponding side-effects (if any).
Let us briefly explain the transition rules:
\begin{itemize}
\item Rule \textit{Seq} considers the reduction of a sequential
expression, which is denoted by a transition labelled with $\tau$
at the expression level. In this case, no side-effect is required. 
Therefore, the rule only updates the process configuration
with the new values, $\theta',e',S'$, and adds 
a new item $\mathsf{seq}(\theta,e,S)$ 
to the history, so that a backward step 
becomes trivially deterministic.
Although this is orthogonal to the topic of this paper, 
the stored information can be optimized, e.g., along 
the lines of \cite{NPV18}.

\item As for sending a message, we distinguish two cases. When the
process log is empty, we tag the message with a fresh identifier; when
the log is not empty, the tag is obtained from an element 
$\mathsf{send}(\ell)$ in the log (which is then removed). 
In both cases, besides adding the new message to the global
mailbox, a new item of the form $\mathsf{send}(\theta,e,S,p',\{v,\ell\})$
is added to the history so that the step becomes reversible.

\item Receiving a message proceeds much in a similar way. 
We also have two rules depending on whether the process log
is empty or not.\footnote{Here, we use the auxiliary function 
$\mathsf{match\_rec}$
to select the matching clause, so that it returns the matching
substitution as well as the body of the selected clause.} 
If there is no log, an arbitrary message is received.
Otherwise, if we have an item $\mathsf{rec}(\ell)$ in the log,
only a message labeled with $\ell$ can be received. The history
is anyway updated with a term of the form
$\mathsf{rec}(\theta,e,S,p',\{v,\k\})$. Observe that $\kappa$
(the \emph{future}) is now bound to the body of the selected
clause. 

\item As for spawning a process, we also distinguish two
cases, depending on the process log. If it is empty, then
a fresh pid is chosen for the new process. Otherwise, the
pid in the process log is used. In both cases, a new term
$\mathsf{spawn}(\theta,e,S,p')$ is added to the history
of the process. Moreover, $\kappa$ is bound to the pid
of the spawned process.

\item Finally, rule \textit{Self} simply binds $\kappa$ with the
pid of the selected process, and adds a new term
$\mathsf{self}(\theta,e,S)$ to the process history.
\end{itemize}
Trivially, when no system log is considered, 
the reversible semantics is a conservative extension
of the standard semantics since we only added some additional
information (the history) but imposed no additional
restriction to perform a reduction step.
Moreover, when a system log is provided, one can easily
prove that the reversible semantics is sound and complete
w.r.t.\ the traced computation.

As for the backward (reversible) semantics, it can be easily
obtained by reversing the rules of 
Figure~\ref{fig:combined-replay-rules-uncontrolled}
and, then, removing all unnecessary conditions in the 
premises. The resulting rules are shown in 
Figure~\ref{fig:combined-rollback-rules-uncontrolled}, where
the auxiliary function $\dom$ returns the variables in
the domain of a substitution.
Note that, in these rules, we \emph{always} take an element from the
history and move the corresponding information (if any) to the
system log. Therefore, once we go backward, forward steps will
be driven by the corresponding log, no matter if we initially
considered the log of a computation or not. 

The reversible semantics is denoted by the relation
$\rightleftharpoons$ which is defined as the union of the forward
and backward transition relations ($\rh\cup\lh$).

The main differences with previous versions of the reversible
semantics are summarized as follows:
\begin{itemize}
\item At the level of expressions, we consider the source language,
Erlang, rather than the intermediate representation, Core Erlang.
Moreover, we also consider higher-order expressions, which
were skipped so far.

\item Regarding the reversible semantics, we keep the same structure
of previous versions but integrate both definitions, the user-driven
reversible semantics of \cite{LNPV18jlamp} and the replay 
reversible semantics of \cite{LPV19}. This simplifies the development
of a debugging tool that integrates both definitions into a single
framework.
\end{itemize}
Since our changes mainly affect the control aspects of the reversible
semantics (and the concurrent actions are the same for both
Erlang and Core Erlang), the properties in  
\cite{LNPV18jlamp,LPV19} carry over easily 
to our new approach. Basically, the following 
properties should also hold in our framework:
\begin{itemize}
\item The so-called \emph{loop lemma}: For every pair of
  systems, $s_1$ and $s_2$, we have $s_1 \rh_{p,r} s_2$ iff
  $s_2 \lh_{p,r} s_1$.
  
\item An essential property of reversible systems, 
\emph{causal consistency}, which is stated as follows: 
Given two coinitial (i.e., starting with the same configuration)
derivations $d_1$ and $d_2$, 
then $d_1\approx d_2$ iff $d_1$ and $d_2$ are cofinal
(i.e., they end with the same configuration).

\item Finally, one could also prove that bugs are preserved under 
the reversible semantics: 
a (faulty) behavior occurs in a traced derivation iff the
replay derivation also exhibits the same \emph{faulty} behavior, hence
replay is correct and complete.
\end{itemize}

%%%%%%%%%%%%%%%%%%%%%%%%%%%%%%%%%%%%%%%%%%%%%%%%%%%%%%%%
\subsection{Controlled Semantics} \label{sec:controlled}

In this section, we introduce a controlled version of the reversible 
semantics. The key idea is that this semantics is driven by the user
requests, e.g., ``go forward until the spawning of process $p$", 
``go backwards until the step immediately 
before message $\ell$ was sent", etc.

Here, we consider that, given a system
$s$, we want to start a forward (resp.\ backward) derivation
until a particular
action $\psi$ is performed (resp.\ undone) on a given process
$p$. We denote such a request with the following notation: 
$\sql s \sqr_\Phi$, where $s$ is a system and $\Phi$ is a sequence of
requests that can be seen as a
stack where the first element is the most recent request.
We formalize the requests as a static stream that is
provided to the calculus but, in practice, the requests are
provided by the user in an interactive way. 
In this paper, we consider the following requests:
\begin{itemize}
\item $\{p,\mathsf{s}\}$: one step backward/forward of process
  $p$;\footnote{The extension to $n$ steps is straightforward. We omit
    it for simplicity.
  }
\item $\{p,\k_\Uparrow\}$: a backward/forward derivation of
  process $p$ up to the sending of the
  message tagged with $\k$;
\item $\{p,\k_\Downarrow\}$: a backward/forward derivation of
  process $p$ up to the reception of the
  message tagged with $\k$;
\item $\{p,\mathsf{sp}_{p'}\}$: a backward/forward derivation
  of process $p$ up to the spawning of the
  process with pid $p'$.
\item $\{p,\mathsf{sp}\}$: a backward derivation of process $p$ up to
  the point immediately after its creation;
\item $\{p,X\}$: a backward derivation of process $p$ up to the introduction of variable
  $X$. 
\end{itemize}
When the request can be either a forward or a backward request, we
use an arrow to indicate the direction. 
E.g., $\{p,\overrightarrow{\mathsf{s}}\}$ requires one step forward,
while $\{p,\overleftarrow{\mathsf{s}}\}$ requires one step backward.
In particular, $\{p,\overleftarrow{\mathsf{sp}}\}$ and 
$\{p,\overleftarrow{X}\}$ have just one version since they always
require a backward computation.

A debugging session can start either with a log (computed using
the tracing semantics or, equivalently, an instrumented source program)
or with an empty log. If the log is not empty, we speak of
\emph{replay} debugging; otherwise, we say that it
is a \emph{user-driven} debugging session. Of course, one can start in
replay mode and, once all the actions of the log are consumed,
switch to the user-driven mode.

The requests above are \emph{satisfied} when a corresponding
uncontrolled transition is performed. This is where the third element
labeling the relations of the reversible semantics in
Figures~\ref{fig:combined-replay-rules-uncontrolled} and
\ref{fig:combined-rollback-rules-uncontrolled} comes into play. This
third element is a set with the requests that are satisfied in the
corresponding step.

\begin{figure}[t]
  \centering
  $
  \begin{array}{c}
    \multicolumn{1}{l}{\mbox{\sc Forward rules:}}\\[1ex]
    
    \displaystyle

    \frac{\Log;\Gamma;\Pi \rh_{p,r,\Psi'}
    \Log';\Gamma';\Pi'~~\wedge~~\psi\in\Psi'}{\sql \Log;\Gamma;
    \Pi \sqr_{\{p,\overrightarrow{\psi}\}\cons\Psi} 
    \rightsquigarrow \sql\Log';\Gamma';\Pi'\sqr_{\Psi}}

    \displaystyle  \hspace{7ex}

    \frac{\Log;\Gamma;\Pi \rh_{p,r,\Psi'}
    \Log';\Gamma';\Pi' ~~\wedge~~ \psi\not\in\Psi'}{\sql\Log;\Gamma;\Pi\sqr_{\{p,\overrightarrow{\psi}\}\cons\Psi}
    \rightsquigarrow \sql\Log';\Gamma';\Pi' \sqr_{\{p,\overrightarrow{\psi}\}\cons\Psi}}\\[4ex]

    \displaystyle

    \frac{\Log[\red{p\mapsto \mathsf{rec}(\k)\cons
    \omega}];\Gamma;\Pi \not\rh_{p,r,\Psi'} ~~\wedge~~ \sender(\Log,\k) = p'}
    {\sql \Log[\red{p\mapsto \mathsf{rec}(\k)\cons
    \omega}];\Gamma;\Pi \sqr_{\{p,\overrightarrow{\psi}\}\cons\Psi} 
    \rightsquigarrow
    \sql \Log[\red{p\mapsto \mathsf{rec}(\k)\cons
    \omega}];\Gamma; \Pi\sqr_{(\{p',\overrightarrow{\k_\Uparrow}\},\{p,\overrightarrow{\psi}\})\cons\Psi}}
      \\[4ex]

    \displaystyle

    \frac{%\Gamma;\Pi \not\rh_{p,r,\Psi'}~~\wedge~~
      \not\exists
    p~\mbox{in}~\Pi~~\wedge~~ 
    \parent(\Log,p)=p'}{\sql\Log,\Gamma;\Pi\sqr_{\{p,\overrightarrow{\psi}\}\cons\Psi}
    \rightsquigarrow \sql\Log,\Gamma;\Pi\sqr_{(\{p', \overrightarrow{\mathsf{sp}_{p}}\},\{p,\overrightarrow{\psi}\})\cons\Psi}}\\[4ex]
 
    \multicolumn{1}{l}{\mbox{\sc Backward rules:}}\\[1ex]

    \displaystyle

    \frac{\Log;\Gamma;\Pi \lh_{p,r,\Psi'}
    \Log';\Gamma';\Pi'~~\wedge~~\psi\in\Psi'}{\sql \Log;\Gamma;\Pi \sqr_{\{p,\overleftarrow{\psi}\}\cons\Psi} 
    \rightsquigarrow \sql \Log';\Gamma';\Pi'\sqr_{\Psi}}

    \displaystyle \hspace{7ex}

    \frac{\Log;\Gamma;\Pi \lh_{p,r,\Psi'}
    \Log';\Gamma';\Pi' ~~\wedge~~ \psi\not\in\Psi'}{\sql \Log;\Gamma;\Pi\sqr_{\{p,\overleftarrow{\psi}\}\cons\Psi}
    \rightsquigarrow \sql \Log';\Gamma';\Pi' \sqr_{\{p,\overleftarrow{\psi}\}\cons\Psi}}\\[4ex]

     \displaystyle

    \frac{\Log;\Gamma; \tuple{p,\blue{\mathsf{send}(\theta,e,S,p',\{v,\k\})\cons
           \h},\theta',e'}\comp \Pi \not\lh_{p,r,\Psi'} }
    {\begin{array}{l}
       \sql \Log;\Gamma; \tuple{p,\blue{\mathsf{send}(\theta,e,S,p',\{v,\k\})\cons
           \h},\theta',e',S'}\comp \Pi \sqr_{\{p,\overleftarrow{\psi}\}\cons\Psi}\\ 
    \hspace{10ex}\rightsquigarrow
    \sql \Log;\Gamma; \tuple{p,\blue{\mathsf{send}(\theta,e,S,p',\{v,\k\})\cons
           \h},\theta',e',S'}\comp \Pi\sqr_{(\{p',\overleftarrow{\k_\Downarrow}\},
       \{p,\overleftarrow{\psi}\})\cons\Psi}
       \end{array}}
      \\[7ex]

    \displaystyle

    \frac{\Log;\Gamma; \tuple{p,
        \blue{\mathsf{spawn}(\theta,e,S,p')\cons\h},\theta',e',S'}
        \comp \Pi \not\lh_{p,r,\Psi'} }
    {\begin{array}{l}
       \sql \Log;\Gamma; \tuple{p,
        \blue{\mathsf{spawn}(\theta,e,S,p')\cons\h},\theta',e',S'}
        \comp \Pi \sqr_{\{p,\overleftarrow{\psi}\}\cons\Psi}\\ 
    \hspace{10ex}\rightsquigarrow
    \sql \Log;\Gamma; \tuple{p,
        \blue{\mathsf{spawn}(\theta,e,S,p')\cons\h},\theta',e',S'}
        \comp \Pi\sqr_{(\{p',\overleftarrow{\mathsf{sp}}\},
       \{p,\overleftarrow{\psi}\})\cons\Psi}
       \end{array}}
      \\[6ex]

    \displaystyle

    \frac{}
    {\sql \Log;\Gamma; \tuple{p,\blue{\nil},\theta',e',S'}\comp
    \Pi \sqr_{\{p,\overleftarrow{\mathsf{sp}}\}\cons\Psi} \rightsquigarrow
    \sql \Log;\Gamma; \tuple{p,\blue{\nil},\theta',e',S'}\comp\Pi\sqr_{\Psi}}
    \\
  \end{array}
  $
  \caption{Controlled forward/backward semantics} \label{fig:controlled-replay-rollback-rules-new}
\end{figure}

Let us explain the rules of the controlled semantics
in Fig.~\ref{fig:controlled-replay-rollback-rules-new}. Here, we assume
that the computation always starts with a single request. We then 
have the following possibilities:
\begin{itemize}
\item If the desired process $p$ can perform a step satisfying
  the request $\psi$ on top of the stack, we do it and remove the request
  from the stack of requests (first rule of both forward and backward
  rules).

\item If the desired process $p$ can perform a step, but the step 
  does not satisfy the request $\psi$, we update the
  system but keep the request in the stack (second rule of both 
  forward and backward rules).

\item If a step on the desired process $p$ is not possible, then we
  track the dependencies and add a new request on top of the 
  stack.\footnote{Note that, if the process' log is empty, only
  the first two rules are applicable; in other words, the user must
  provide feasible requests to drive the forward computation.}
  For the forward rules, either we
  cannot proceed because we aim at receiving a message which
  is not in $\Gamma$ or because the considered process does
  not exist. In the first case, the label $\k$ of the message can be
  found in the process' log. Then, the auxiliary
  function $\sender$ is used to locate the process $p'$ 
  that should send message $\k$, so that an additional
  request for process $p'$ to send message $\k$ is added.  
  In the second case, if process $p$ is not in $\Pi$,
  then we add another request for the parent of $p$ to
  spawn it. For this purpose, we use the auxiliary function
  $\parent$. 
  \\
  
  As for the backward rules, we consider three cases: one rule to add 
  a request to undo the receiving of a message whose sending 
  we want to undo, another rule to undo the actions of a given process 
  whose spawning we want to undo, 
  and a final rule to check that a process has reached its initial state
  (with an empty history), and the request $\{p,\overleftarrow{\mathsf{sp}}\}$ can be
  removed. In this last case, the process $p$ will actually be removed
  from the system when a request of the form $\{p',\overleftarrow{\mathsf{sp}_{p}}\}$ is on top
  of the stack.
\end{itemize}
The relation ``$\rightsquigarrow$" can be seen as a controlled version of
the uncontrolled reversible semantics ($\rightleftharpoons$) 
in the sense that each
derivation of the controlled semantics corresponds to a derivation of
the uncontrolled one, while the opposite is not generally true.

\begin{figure}[h]
    \centering
    \includegraphics[width=0.52\textwidth]{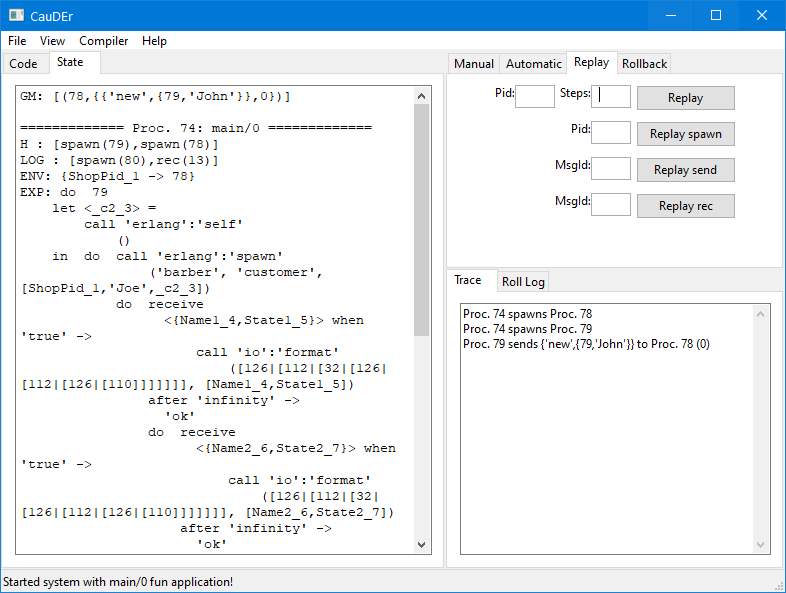}~~
    \includegraphics[width=0.48\textwidth]{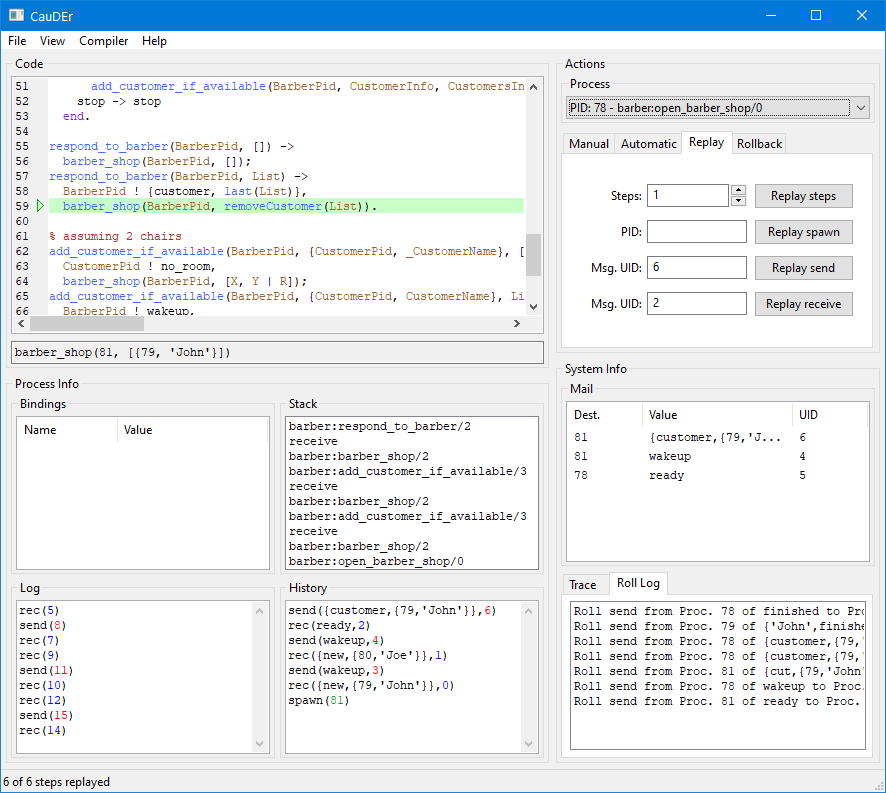}
    \caption{CauDEr Old User Interface vs New Interface}
    \vspace{-2ex}
    \label{fig:cauder-ui}
\end{figure}

The controlled semantics is the basis of the implemented 
reversible debugger \textsf{CauDEr}. Figure~\ref{fig:cauder-ui}
shows a snapshot of both the old and the 
new, improved user interface. In contrast to the previous
version of the debugger \cite{Cauder}, we show the source 
code of the program and \emph{highlight} 
the line that is being evaluated.
In contrast, the old version showed the current Core Erlang
expression to be reduced, which was far less intuitive. 
Moreover, all the available information (bindings, stack, 
log, history, etc) is shown in different boxes, while the previous
version showed all information together in a single text box.
Furthermore, the user can now decide whether to add a log or not,
while the previous version required the use of 
different implementations of the debugger.

The new version of the reversible debugger is publicly available from
\url{https://github.com/mistupv/cauder}.

%%%%%%%%%%%%%%%%%%%%%%%%%%%%%%%%%%%%%%%%%%%%%%%%%%%%%%%%%%
\section{Conclusions and Future Work}\label{sec:concl} 

In this paper, we have adapted and extended the framework
for causal-consistent reversible debugging from Core Erlang
to Erlang. In doing so, we have extended the standard
semantics to also cope with program constructs that were
not covered in previous work \cite{LNPV18jlamp,LNPV18,LPV19},
e.g., \emph{if} statements, higher-order functions, sequences, etc.
Furthermore, we have integrated user-driven debugging
and replay debugging into a single, more general framework.
Finally, the user interface of \textsf{CauDEr} has been 
redesigned in order to make it easier to use (and closer to
that of the standard debugger of Erlang). 
We refer the reader to \cite{LNPV18jlamp,LPV19} for
a detailed comparison between causal-consistent reversible
debugging in Erlang and other, related work.

As for future work, we aim at modelling in the semantics 
different levels of granularity
for debugging. For instance, the user may choose to 
evaluate a function call in one step or to explore the reduction
of the function's body step by step. Moreover, we are also
exploring the possibility of allowing the user to 
\emph{speculatively} receive a message which is different from
the one in the process' log during replay debugging.
Finally, other interesting ideas for future work include
the implementation of appropriate extensions to deal with 
distributed programs and error handling (following, e.g.,
the approach of \cite{LM20}).

\subsubsection*{Acknowledgements.}

The authors would like to thank Ivan Lanese for his useful remarks that 
helped us to improve the new version of the CauDEr debugger.

\bibliographystyle{splncs03}
%\bibliography{biblio}

%%%%%%%%%%%%%%%%%%%%%%%%%%%%%%%%%%%%%%%%%%%%%%%%%%%%%%%%%%
%%%%%%%%%%%%%%%%%%%%%%%%%%%%%%%%%%%%%%%%%%%%%%%%%%%%%%%%%%
\clearpage
\appendix

\section{The Language Syntax}

In this section, we present the complete syntax of the 
considered language: a significant
subset of the higher-order functional and concurrent programming 
language Erlang.

\begin{figure}[t]
\centering
  $
  \begin{array}{rcl@{~~~}rcl}
    \mathit{program} & ::= & 
    \mathit{mod}_1~\ldots~\mathit{mod}_n \\
    \mathit{mod} & ::= &
                             \mathit{fun\_def}_1~\ldots~\mathit{fun\_def}_n \\
    \mathit{fun\_def} & ::= & \mathit{fun\_rule} ~\{';'~\mathit{fun\_rule}\} ~'.'\\
    {\mathit{fun\_rule}} & ::= & \mathit{Atom}([\mathit{exprs}])~ [\mathsf{when}~ \mathit{guard}] \to \mathit{exprs} \\
    \mathit{fun\_expr} & ::= & \mathsf{fun}~\mathit{fun} ~\{';'~\mathit{fun}\} ~\mathsf{end}\\
    {\mathit{fun}} & ::= & ([\mathit{exprs}])~ [\mathsf{when}~ \mathit{guard}] \to \mathit{exprs} \\
    %\mathit{guard} & ::= & \mathit{exprs}~ \{';'~\mathit{exprs}\}\\
    \mathit{pattern} & ::= 
    & \mathit{atomic} \mid \mathit{Var} \mid~ '\{' [\mathit{patterns}]
                          '\}' \mid~ '[' ~[\mathit{patterns} ~'|'
                          \mathit{pattern} ]~ ']' \\
    \mathit{patterns} & ::= & \mathit{pattern}~\{','~\mathit{pattern}\}\\ 
    \mathit{exprs} & ::= & \mathit{expr}~\{','~\mathit{expr}\}\\ 
    \mathit{expr} & ::= & \mathit{atomic} \mid \mathit{Var} \mid~ '\{' [\mathit{exprs}]
                          '\}' \mid~ '[' ~[\mathit{exprs} ~'|'
                          \mathit{expr} ]~ ']' \mid
                          \mathsf{if} ~\mathit{if\_clauses}
                          ~\mathsf{end}\\
    & \mid & \mathsf{case}~ expr ~\mathsf{of}~
             \mathit{cr\_clauses}~\mathsf{end} \mid
             \mathsf{receive}~ \mathit{cr\_clauses} ~\mathsf{end} \mid
             \mathit{expr}~!~\mathit{expr}\\
    & \mid & \mathit{pattern} = \mathit{expr} \mid 
    [\mathit{Mod}\! :]\mathit{expr}([\mathit{exprs}]) \mid
    \mathit{fun\_expr} \mid Op exprs\\
    \mathit{atomic} & ::= & \mathit{Atom} \mid \mathit{Char} \mid
                            \mathit{Float} \mid \mathit{Integer} \mid
                            \mathit{String}\\
    \mathit{if\_clauses} & ::= & \mathit{guard}~\to~\mathit{exprs}
                                 ~\{';'~
                                 \mathit{guard}~\to~\mathit{exprs}\}\\
    \mathit{cr\_clauses} & ::= & \mathit{pattern}~ [\mathsf{when}~
                                 \mathit{guard}] \to \mathit{exprs}
                                 ~\{';'~\mathit{pattern}~ [\mathsf{when}~
                                 \mathit{guard}] \to
                                 \mathit{exprs}\}\\
    
  \end{array}
  $
\caption{Language syntax rules} \label{fig:syntax}
\end{figure}

The syntax of the language is shown in Figure~\ref{fig:syntax}.
Terminals are denoted with sans serif font or using single quotes.
We distinguish \emph{expressions}, \emph{patterns}, and \emph{values}.
Patterns are built from variables,
literals (atomic values), lists, and tuples; they can only contain fresh
variables.  In contrast, values are built from literals, lists, and
tuples, i.e., they are \emph{ground} (without variables) patterns.
Expressions are ranged over by $e,e',e_1,\ldots$,
patterns by $pat$, $pat'$, $pat_1$, $\ldots$ and values by
$v,v',v_1,\ldots$ Atoms (i.e., constants with a name) are written in
roman letters, while variables start with an uppercase letter.

A program is a collection of modules, where each module
includes a sequence of function definitions,
%\footnote{For simplicity, we do not consider modules in this work.} 
A function is defined by one or more clauses of the 
form $f(pat_1,\ldots,pat_n) \to e_1,\ldots,e_m$, where $f$ is an atom.
Note that clauses can have a sequence of expressions 
in the right-hand side. Clauses are tried from top to
bottom using pattern matching. 
Morever, clauses may have a \emph{guard} 
(prefixed by the keyword \textsf{when}) that must be
evaluated to $\mathsf{true}$ in order for the rule to be
applicable. Guards can only contain calls to predefined functions
(typically, relational and arithmetic operators). 
%In the paper, we
%consider that a program consists of a single module for simplicity.
%The body of a function is a sequence of expressions, where each
An expression can include atomic values, variables, tuples of the form
$\{e_1,\ldots,e_n\}$, lists (using Prolog-like notation, where
$\nil$ is the empty list and
$[e_1|e_2]$ denotes a list with head $e_1$ and tail
$e_2$), conditionals, case expressions, receive expressions, messsage
sending, pattern matching (i.e., expressions of the form
$pat = expr$), function applications, 
anonymous functions, and the usual arithmetic and relational operators. 
As in Erlang, the only
data constructors in the language (besides literals) are the
predefined functions for lists and tuples.

Erlang includes a number of built-in functions (BIFs). In this work,
we only consider \textsf{self}/0, which returns the process identifier
of the current process, and \textsf{spawn}/1 and \textsf{spawn/3}
that creates a new process (see below).

%%%%%%%%%%%%%%%%%%%%%%%%%%%%%%%%%%%%%%%%%
\section{The Language Semantics}

Now, we present the semantics of the language.
As in \cite{LNPV18jlamp}, our operational semantics is defined by
means of several layers. The first, lower level layer considers
expressions. Here, we distinguish sequential expressions from 
those related to concurrency. The sequential subset of the language
is a typical higher-order eager functional programming language.
First, we consider the semantics of sequential 
expressions. 

\subsection{Sequential Expressions}

We need some preliminary notions.
A \emph{substitution} $\theta$ is a mapping
from variables to expressions, and $\dom(\theta) =
\{X\in\mathit{Var} \mid X \neq \theta(X)\}$ is its
domain. Substitutions are usually denoted by (finite) sets of bindings like,
e.g., $\{X_1\mapsto v_1,\ldots,X_n\mapsto v_n\}$.
Substitutions are extended to morphisms from expressions to
expressions in the natural way.  The identity substitution is denoted
by $\id$. Composition of substitutions is denoted by juxtaposition,
i.e., $\theta\theta'$ denotes a substitution $\theta''$ such that
$\theta''(X) = \theta'(\theta(X))$ for all $X\in\mathit{Var}$. We
follow a postfix notation for substitution application: given an
expression $e$ and a substitution $\sigma$ the application $\sigma(e)$
is denoted by $e\sigma$.

We often denote by $\ol{o_n}$ a sequence of syntactic objects
$o_1,\ldots,o_n$ for some $n$. We use $\ol{o}$ when the number 
of objects is not relevant.

In the following, we consider that each expression can be written as
$C[e]$ where $e$ is the next \emph{redex} to be reduced according to
the usual eager semantics. E.g., an expression like
``$\tt \mathsf{case}~f(42)~\mathsf{of}~\ldots~\mathsf{end}$'' can be
represented as $C[{\tt f(42)}]$ since the call $\tt f(42)$ is needed to evaluate
the case expression.

\begin{figure}[t]
  \[
    \begin{array}{c}
    (\mathit{Var}) ~ {\displaystyle \frac{}{\theta,C[X],S
    \arro{\tau} \theta,C[\theta(X)],S}}

    \\[4ex]

      (\mathit{Seq1}) ~ {\displaystyle
        \frac{}{\theta,C[v,e],S
        \arro{\tau} \theta,C[e],S}}  ~~~~
      
    (\mathit{Seq2}) ~ {\displaystyle
        \frac{~}{\theta,v,\mathsf{seq}(C[\_])\consl S
        \arro{\tau} \theta,C[v],S}} \\[4ex]

      (\mathit{If}) ~ {\displaystyle
        \frac{\mathsf{eval\_guard}(g_1\theta,\ldots,g_n\theta) = i
    }{\theta,C[\mathsf{if}~g_1\to
    e_1;\ldots;g_n\to e_n~\mathsf{end}],S
        \arro{\tau} \theta,e_i,\mathsf{seq}(C[\_])\consl S}} \\[4ex]

    (\mathit{Case}) ~ {\displaystyle
        \frac{\mathsf{match\_case}(v,cl_1\theta,\ldots,cl_n\theta) = \tuple{\theta_i,e_i}}{\theta,C[\mathsf{case}~v~\mathsf{of}~cl_1;\ldots;cl_n~\mathsf{end}],S
        \arro{\tau} \theta\theta_i,e_i,\mathsf{seq}(C[\_])\consl S}} \\[4ex]

     (\mathit{Match}) ~ {\displaystyle
        \frac{\mathsf{match}(pat\theta,v) = \sigma}{\theta,C[pat = v],S
        \arro{\tau} \theta\sigma,C[v],S}} \\[4ex]

    (\mathit{Op}) ~ {\displaystyle
      \frac{\mathsf{eval}(op,v_1,\ldots,v_n)=v}{\theta,C[op(v_1,\ldots,v_n)],S
        \arro{\tau} \theta,C[v],S}} \\[4ex]

      (\mathit{Fun}) ~ {\displaystyle
      \frac{}{\theta,C[\mathsf{fun}~\mathit{fun_1};\ldots;\mathit{fun_m} ~\mathsf{end}],S
        \arro{\tau} \theta,C[\tuple{\theta, \mathsf{fun}~\mathit{fun_1};\ldots;\mathit{fun_m} ~\mathsf{end}}],S}} \\[4ex]

    (\mathit{Call1}) ~ {\displaystyle
    \frac{\mathsf{match\_fun}((v_1,\ldots,v_n),\mathsf{def}(f/n,\mathsf{P})) = (\sigma,e)}{\theta,C[f(v_1,\ldots,v_n)],S
        \arro{\tau} \sigma,e,(\theta,C[\_])\consl S}} %\{\ol{X_n\mapsto v_n}\},e}}
    \\[4ex]

    (\mathit{Call2}) ~ {\displaystyle
    \frac{\mathsf{match\_fun}((v_1,\ldots,v_n), \tuple{\theta', \mathsf{fun}~\mathit{fun_1};\ldots;\mathit{fun_n} ~\mathsf{end}}) = (\sigma,e)}{\theta,C[\tuple{\theta', \mathsf{fun}~\mathit{fun_1};\ldots;\mathit{fun_m} ~\mathsf{end}} \:(v_1,\ldots,v_n)],S
        \arro{\tau} \sigma,e,(\theta,C[\_])\consl S}} %\{\ol{X_n\mapsto v_n}\},e}}
    \\[4ex]

        (\mathit{Return}) ~ {\displaystyle
      \frac{}{\sigma,v,(\theta,C[\_])\consl S
        \arro{\tau} \theta,C[v],S}} %\{\ol{X_n\mapsto v_n}\},e}} 

  \end{array}
  \]
\caption{Standard semantics: evaluation of sequential expressions} \label{fig:seq-rules}
\end{figure}

The rules for sequential expressions are shown in Figure~\ref{fig:seq-rules}.
Our (labeled) reduction semantics is defined on triples $\theta,e,S$, where
$\theta$ is the current environment that stores the values of the program
variables, $e$ is the expression to be evaluated, and $S$ is a stack.
Let us briefly explain the rules of our semantics:
\begin{itemize}
\item Rule \textit{Var} evaluates a variable by simply looking up his value
in the environment.
\item Rule \textit{Seq1} removes a value from a sequence, so that the 
remaining elements can then be reduced. 
In general, several Erlang expressions can be reduced to a sequence.
Unfortunately, this may give rise to an expression which is syntactically
ilegal. Consider, e.g., that the evaluation of the above expression
$\tt \mathsf{case}~f(42)~\mathsf{of}~\ldots~\mathsf{end}$ reduces
$\tt f(42)$ to a sequence like $X=42,X$. Then, we would get the
expression $\tt \mathsf{case}~X=42,X~\mathsf{of}~\ldots~\mathsf{end}$,
which is not syntactically correct (the argument of a case expression
must be a single expression, sequences are not allowed).
To overcome this problem, rules \textit{If} and \textit{Case} move
the current context to the stack until the reduced expression is
eventually reduced to a single expression and the
context is recovered by rule \textit{Seq2}.

\item Rule \textit{If} evaluates the guards using the auxiliary function
$\mathsf{eval\_guard}$, which returns the index of the first guard
that evaluates to true. As mentioned above, we move the current 
context to the stack. 

\item Rule \textit{Case} matches the case argument, $v$ with the
case branches (possibly including guards) using the auxiliary
function $\mathsf{match\_case}$ and returns a pair
$\tuple{\theta_i,e_i}$ with the matching substitution and the 
expression in the selected branch. As before, the context
is moved to the stack to avoid giving rise to an illegal expression
if $e_i$ is a sequence.

\item Rule \textit{Match} evaluates a pattern matching equation using
auxiliary function $\mathsf{match}$, which returns the matching
substitution.

\item Rule \textit{Op} evaluates arithmetic and relational operations
using the auxiliary function $\mathsf{eval}$.

\item Rule \textit{Fun} evaluates an anonymous function by reducing
it to a closure.

\item Rules \textit{Call1} and \textit{Call2} evaluate a function application
by moving the current context and environment to the stack and then
reducing the function's body using the auxiliary function
\textsf{match\_fun} that takes the arguments of the function call
and either the definition of the given function in the source program
or a closure. Function \textit{Return} recovers the context and
environment from the stack once the function's body is reduced
to a value.
\end{itemize}

%%%%%%%%%%%%%%%%%%%%%%%%%%%%%%%%%%%%%%%%%
\subsection{Concurrency}

In this section, we consider the semantics of concurrent actions. 
Essentially, a running application consists of a number of processes
that interact by sending and receiving messages. Message sending
is asynchronous, while receiving messages may block a process
if no (matching) message arrived yet. Processes are uniquely
identified by their \emph{pid} (process identifier).

In principle, one can consider that each process has an associated
mailbox or queue where incoming messages are stored until they are
consumed by a receive expression. 
When a process sends a message, it is eventually, 
stored in the mailbox of the target process.\footnote{In this work, we
do not consider lost messages.} Between the sending of a message
and storing it in a process' mailbox, we say that the message is 
in the \textit{network} (which is called the \textit{ether} in \cite{SFB10}).

In this work, similarly to \cite{LPV19}, our
semantics abstracts away from processes' queues. Here, we only
consider a single data structure, called \emph{global mailbox},
that represents both the network and the processes' mailboxes. 
Furthermore, the semantics represents an overapproximation
of the actual semantics of Erlang since we impose no restriction on 
the order of messages. In Erlang, the messages between two given
processes must arrive in the same order they were sent. We skip
this restriction for simplicity (but could easily be implemented, 
see \cite{NPV16b}).
Nevertheless, removing this restriction is not relevant in
\emph{replay} mode, since the
debugger will follow the trace of an actual execution in some Erlang
environment and, thus, only executions that respect the above
restriction can be considered.  

We note that this contrasts with previous semantics, e.g.,\cite{LNPV18jlamp,NPV16b,SFB10}, where both the network
and the processes' mailboxes were explicitly modeled.

Let us now introduce the notions of \emph{process} and \emph{system},
which are essential in our semantics.

\begin{definition}[process]
  A process is denoted by a configuration of the form $\tuple{p,\theta,e,S}$,
  where 
  $p$ is the pid of the process, 
  $\theta$ is an environment (a substitution of values for variables),
  $e$ is an expression to be evaluated, and 
  $S$ is an stack.
\end{definition}
A so called \emph{global mailbox} is  then used to store
sent messages until they are delivered (i.e., \emph{consumed}
by a process using a receive expression):

\begin{definition}[global mailbox]
  We define a global mailbox, $\Gamma$, as a multiset of triples of the
  form \linebreak $(sender\_pid,target\_pid,message)$.  Given a global mailbox
  $\Gamma$, we let $\Gamma\cup\{(p,p',v)\}$ denote a new mailbox also
  including the triple $(p,p',v)$, where we use ``$\:\cup$'' as multiset
  union.
\end{definition}
Finally, a \emph{system} is defined as follows:

\begin{definition}[system] \label{def:newsystem} A system is a pair
  $\Gamma;\Pi$, where $\Gamma$ is a global
  mailbox % }, is a data structure modeling message communication,
  and $\Pi$ is a pool of processes, denoted as %by an expression of the
  %form
  $ \tuple{p_1,\theta_1,e_1,S_1} ~\comp \cdots
  \comp~\tuple{p_n,\theta_n,e_n,S_n} $; here ``$\comp\!$'' represents an
  associative and commutative operator.
  We often denote a system as
  $ \Gamma; \tuple{p,\theta,e,S}\comp\Pi $ to point out that
  $\tuple{p,\theta,e,S}$ is an arbitrary process of the pool
  (thanks to the fact that ``$\comp$'' is associative and
  commutative).

  A \emph{initial system} has the form
  $\{\:\};\tuple{p,\id,e,\nil}$, where $\{\:\}$ is an empty global mailbox,
  $p$ is a pid, $\id$ is the identity substitution, $e$ is an
  expression (typically a function application that starts the
  execution), and $\nil$ is an empty stack.
\end{definition}

\begin{figure}[t]
  \[
  \begin{array}{r@{~}c}
      (\mathit{SendExp}) & {\displaystyle
          \frac{}{\theta,C[v_1\:!\: v_2],S \arro{\mathsf{send}(v_1,v_2)} \theta,C[v_2],S}
          }\\[4ex]

    (\mathit{ReceiveExp}) & {\displaystyle
      \frac{}{\theta,C[\mathsf{receive}~cl_1;\ldots;cl_n~\mathsf{end}],S
        \arro{\mathsf{rec}(\kappa,\ol{cl_n})}
        \theta,\kappa,\mathsf{seq}(C[\_])\consl S
      }
      }\\[4ex]

     (\mathit{SpawnExp1}) & {\displaystyle
       \frac{}{\theta,C[\mathsf{spawn}(mod,f,[\ol{v_n}])],S
         \arro{\mathsf{spawn}(\kappa,mod,f,[\ol{v_n}])} \theta,C[\kappa],S         
       }}\\[4ex]

     (\mathit{SpawnExp2}) & {\displaystyle
       \frac{}{\theta,C[\mathsf{spawn}(\mathsf{fun}()
                         \to \mathit{exprs}~\mathsf{end})],S
         \arro{\mathsf{spawn}(\kappa,\mathit{exprs})} \theta,C[\kappa],S         
       }}\\[4ex]

    (\mathit{SelfExp}) & {\displaystyle
     \frac{}{\theta,C[\mathsf{self}()],S \arro{\mathsf{self}(\kappa)} \theta,C[\kappa],S}}
  \end{array}
  \]
\caption{Standard semantics: evaluation expressions with side-effects} \label{fig:concurrent-rules}
\end{figure}

\begin{figure}[t]
\centering
  $
  \begin{array}{r@{~~}c}
    (\mathit{Seq}) & {\displaystyle
      \frac{\theta,e,S\arro{\tau} \theta',e',S'
      }{\Gamma;\tuple{p,\theta,e,S}\comp \Pi \hoo_{\red{p,\mathsf{seq}}}
      \Gamma;\tuple{p,\theta',e',S'}\comp \Pi}
      }\\[3ex]

    (\mathit{Send}) & {\displaystyle
      \frac{\theta,e,S \arro{\mathsf{send}(p',v)}
        \theta',e',S' ~\mbox{and}~\red{\k~\mbox{is a fresh symbol}}}{\Gamma;\tuple{p,\theta,e,S} 
        \comp \Pi 
                      %\;\instarro{\red{\{p,\mathsf{send}(\k)\}}}\; 
       \hoo_{\red{p,\mathsf{send}(\k)}}
       %\ins{\Gamma}{p}{p'}{\red{\{v,\k\}}};
                      \Gamma\cup\{(p,p',\red{\{v,\k\}})\};\tuple{p,\theta',e',S'}\comp \Pi}
      }\\[3ex]

      (\mathit{Receive}) & {\displaystyle
        \frac{\theta,e,S \arro{\mathsf{rec}(\kappa,\ol{cl_n})}
          \theta',e',S'~~\mbox{and}~~ \mathsf{match\_rec}(\ol{cl_n}\theta,v) =
         (\theta_i,e_i)} {\Gamma\cup\{(p',p,\red{\{v,\k\}})\};
         % \ext{\Gamma}{p'}{p}{\red{\{v,\k\}}};
           \tuple{p,\theta,e,S}\comp
                           \Pi  %%\;\instarro{\red{\{p,\mathsf{rec}(\k)\}}}\;
          \hoo_{\red{p,\mathsf{rec}(\k)}}
           \Gamma;\tuple{p,\theta'\theta_i,e'\{\kappa\mapsto
            e_i\},S'}\comp \Pi}
      }\\[3ex]  

      (\mathit{Spawn1}) & {\displaystyle
        \frac{\theta,e,S \arro{\mathsf{spawn}(\kappa,mod,f,[\ol{v_n}])}
          \theta',e',S'~~\mbox{and}~~ p'~\mbox{is a fresh pid}}{\Gamma;\tuple{p,\theta,e,S} 
          \comp \Pi %%\;\instarro{\red{\{p,\mathsf{spawn}(p')\}}}\; 
           \hoo_{\red{p,\mathsf{spawn}(p')}}
                          \Gamma;\tuple{p,\theta',e'\{\kappa\mapsto
                          p'\},S'}\comp \tuple{p',\id,mod\consl f(\ol{v_n}),\nil} 
          \comp \Pi}
      }\\[3ex]

      (\mathit{Spawn2}) & {\displaystyle
        \frac{\theta,e,S \arro{\mathsf{spawn}(\kappa,\mathit{exprs}])}
          \theta',e',S'~~\mbox{and}~~ p'~\mbox{is a fresh pid}}{\Gamma;\tuple{p,\theta,e,S} 
          \comp \Pi %%\;\instarro{\red{\{p,\mathsf{spawn}(p')\}}}\; 
           \hoo_{\red{p,\mathsf{spawn}(p')}}
                          \Gamma;\tuple{p,\theta',e'\{\kappa\mapsto
                          p'\},S'}\comp \tuple{p',\id,\mathit{exprs},\nil} 
          \comp \Pi}
      }\\[3ex]

    (\mathit{Self}) & {\displaystyle
      \frac{\theta,e,S \arro{\mathsf{self}(\kappa)} \theta',e',S'}{\Gamma;\tuple{p,\theta,e,S} 
        \comp \Pi \hoo_{\red{p,\mathsf{self}}} \Gamma;\tuple{p,\theta',e'\{\kappa\mapsto p\},S'} 
        \comp \Pi }
      }
  \end{array}
  $
\caption{Tracing semantics} \label{fig:tracing-semantics}
\end{figure}

The transition rules for concurrent expressions is shown in
Figure~\ref{fig:concurrent-rules}, while the (labeled) transition rules for
systems is shown in Figure~\ref{fig:tracing-semantics}. 
For the moment, the reader can safely ignore the labels
of the arrows. 
Let us briefly explain how concurrent expressions and systems
are evaluated:
\begin{itemize}
\item First, a system in which the selected process has a sequential 
expression (i.e., an expression that can be reduced using the rules
of Figure~\ref{fig:seq-rules}) is evaluated using rule \textit{Seq} in
Figure~\ref{fig:tracing-semantics} in the obvious way.

\item \textit{Sending a message}. On the one hand, 
rule \textit{SendExp} reduces an expression $v_1 \:!\: v_2$ 
(i.e., sending message $v_2$ to process with pid $v_1$) to 
$v_2$. However, we also need some side-effect: message $v_2$
must be eventually received by process $v_1$. Since this is not
observable locally, we label the step with $\mathsf{send}(v_1,v_2)$
so that rule \textit{Send} can take care of this by adding the triple
$(p,p',\{v,l\})$ to the global mailbox $\Gamma$, where $p$
is the pid of the sender, $p'$ is the pid of the target, and
$\{v,\ell\}$ is a \emph{tagged} message. Here, messages are
tagged with unique labels so that we can
connect messages sent and received without ambiguity. Note that
without unique labels, messages with the same value would be
indistinguishable.

\item \textit{Receiving a message}. At the level of expressions, rule
\textit{ReceiveExp} returns a fresh variable, $\kappa$, since the
receive expression cannot be reduced at this level without accessing
to the global mailbox. Here, $\kappa$ can be seen as a \emph{future}
that will be bound in the next layer of the semantics. As before, we
label the step with enough information for rule \textit{Receive} to
complete the reduction. This rule looks for some (in principle, arbitrary)
message addressed to the considered process in the global mailbox,
checks that it matches some branch of the receive expression using
the auxiliary function $\sf match\_rec$, and then proceeds as in the
evaluation of a case expression. The main difference is that, now, 
$\kappa$ is bound to the selected branch and the message is
removed from the global mailbox.

\item \textit{Spawning a process}. For spawning a process, we proceed
analogously to the previous case. Rules \textit{SpawnExp1} and 
\textit{SpawnExp2} labels the step with the appropriate information
and the calls are reduced to a fresh variables $\kappa$. Then,
rules \textit{Spawn1} and \textit{Spawn2} perform the corresponding
side effect (creating a new process) and bind $\kappa$ to the pid
of the new process. 

\item Finally, $\mathsf{self}()$ returns the pid of the current process
using rule \textit{SelfExp} and \textit{Self}, analogously to the
previous cases.
\end{itemize}
We  refer to reduction steps derived by the system rules as
\emph{actions} taken by the chosen process.

Finally, by considering the labels in the transition steps of the
semantics shown in Figure~\ref{fig:tracing-semantics}, we
get a \emph{tracing} semantics that produces the log of
a computation. This log can then be used in the debugger
to replay a (typically faulty) computation. Let us note that,
in practice, we use a program instrumentation rather than
an instrumented semantics, so that the instrumented program
can be executed in the standard Erlang/OTP environment.
The equivalence between the two alternatives is rather
straightforward though.

In the following, (ordered) sequences are denoted by
$w = (r_1,r_2,\ldots,r_n)$, $n\geq 1$, where $\nil$ denotes the empty
sequence. Given sequences $w_1$ and $w_2$, we denote their
concatenation by $w_1\cons w_2$; when $w_1$ just contains one
element, i.e., $w_1=(r)$, we write $r\cons w_2$ instead of
$(r)\cons w_2$ for simplicity. 

\begin{definition}[log] \label{def:trace} A \emph{log} is a (finite)
  sequence of events $(r_1,r_2,\ldots)$ where each $r_i$ is either
  $\mathsf{spawn}(p)$, $\mathsf{send}(\k)$ or $\mathsf{rec}(\k)$,
  with $p$ a pid and $\k$ a message identifier. Logs are ranged over
  by $\omega$.
  Given a derivation $d = (s_0 \hoo_{p_1,r_1} s_1 \hoo_{p_2,r_2}
  \ldots \hoo_{p_n,r_n} s_n)$,
  $n\geq 0$,
  under the logging semantics, the \emph{log of a process $p$ in $d$}, in
  symbols $\trace(d,p)$, is inductively defined as follows:
  \[
    \trace(d,p) =
    \left\{\begin{array}{l@{~~~}l}
             \nil & \mbox{if}~n=0~\mbox{or}~p~\mbox{does not occur in}~d \\
             r_1\cons\trace(s_1 \hoo^\ast s_n,p) & \mbox{if}~n>0,~p_1=p,~\mbox{and}~r_1\not\in\{\mathsf{seq},\mathsf{self}\}\\
             \trace(s_1 \hoo^\ast s_n,p) & \mbox{otherwise}\\
         \end{array}\right.
 \]
 The \emph{log of $d$}, written $\trace(d)$, is defined as:
 $\trace(d) = \{ (p,\trace(d,p))\mid \mbox{$p$ occurs in
   $d$}\}$.
  We sometimes call $\trace(d)$ the \emph{system} log of $d$ to avoid
  confusion with $\trace(d,p)$ (the process' log).
  Trivially, $\trace(d,p)$ can be obtained from $\trace(d)$, i.e.,
  $\trace(d,p) = \omega$ if $(p,\omega)\in\trace(d)$ and
  $\trace(d,p)=\nil$ otherwise.
\end{definition}
Clearly, given a finite derivation $d$, the associated log $\trace(d)$
is finite too. However, the opposite is not true: we might have a
finite log associated to an infinite derivation (e.g., by applying
infinitely many times rule $\mathit{Seq}$).

\end{document}